\documentclass[aps,prl,reprint,amsmath,amssymb,showpacs,nobibnotes]{revtex4-1}
\usepackage{graphicx}
\usepackage{hyperref} 
\usepackage{dsfont}    
\usepackage{color}

\newcommand\trick[1]{}

\newcommand{\be}{\begin{equation}}
\newcommand{\ee}{\end{equation}}
\newcommand{\eq}[1]{(\ref{#1})}

\newcommand{\bit}{\begin{itemize}}  \newcommand{\eit}{\end{itemize}}
\newcommand{\ben}{\begin{enumerate}}  \newcommand{\een}{\end{enumerate}}

\newcommand{\rf}[1]{(\ref{#1})}

\def\bd{\begin{document}}
\def\ed{\end{document}}
\def\bea{\begin{eqnarray}}
\def\eea{\end{eqnarray}}
\let\bm=\bibitem

\def\la{\langle}
\def\ra{\rangle}

\def\npb#1#2#3{Nucl. Phys. {\bf{B#1}} #3 (#2)}
\def\plb#1#2#3{Phys. Lett. {\bf{#1B}} #3 (#2)}
\def\prl#1#2#3{Phys. Rev. Lett. {\bf{#1}} #3 (#2)}
\def\prd#1#2#3{Phys. Rev. {D bf{#1}} #3 (#2)}
\def\cmp#1#2#3{Comm. Math. Phys. {\bf{#1}} #3 (#2)}
\def\cqg#1#2#3{Class. Quantum Grav. {\bf{#1}} #3 (#2)}
\def\nppsa#1#2#3{Nucl. Phys. B (Proc. Suppl.) {\bf{#1A}}#3 (#2)}
\def\ap#1#2#3{Ann. of Phys. {\bf{#1}} #3 (#2)}
\def\ijmp#1#2#3{Int. J. Mod. Phys. {\bf{A#1}} #3 (#2)}
\def\rmp#1#2#3{Rev. Mod. Phys. {\bf{#1}} #3 (#2)}
\def\mpla#1#2#3{Mod. Phys. Lett. {\bf A#1} #3 (#2)}
\def\jhep#1#2#3{J. High Energy Phys. {\bf #1} #3 (#2)}
\def\atmp#1#2#3{Adv. Theor. Math. Phys. {\bf #1} #3 (#2)}

\def\N{{\cal N}}
\def\sst{\scriptscriptstyle}
\def\thetabar{\bar\theta}
\def\Tr{{\rm Tr}}
\def\one{\mbox{1 \kern-.59em {\rm l}}}

\def\a{\alpha}      \def\da{{\dot\alpha}}  \def\dA{{\dot A}}
\def\b{\beta}       \def\db{{\dot\beta}}
\def\g{\gamma}  \def\G{\Gamma}  \def\dc{{\dot\gamma}}
\def\d{\delta}  \def\D{\Delta}  \def\ddt{\dot\delta}
\def\e{\epsilon}
\def\ve{\varepsilon}
\def\uve{\upvarepsilon}
\def\f{\phi}    \def\F{\Phi}    \def\vvf{\f}
\def\vphi{\varphi}
\def\h{\eta}
\def\k{\kappa}
\def\l{\lambda} \def\L{\Lambda}
\def\m{\mu} \def\n{\nu}
\def\o{\omega}
\def\p{\pi} \def\P{\Pi}
\def\r{\rho}
\def\s{\sigma}  \def\S{\Sigma}
\def\t{\tau}
\def\th{\theta} \def\Th{\Theta} \def\vth{\vartheta}
\def\X{\Xeta}
\def\z{\zeta}

\def\na{\nabla}

\def\cA{{\cal A}} \def\cB{{\cal B}} \def\cC{{\cal C}}
\def\cD{{\cal D}} \def\cE{{\cal E}} \def\cF{{\cal F}}
\def\cG{{\cal G}} \def\cH{{\cal H}} \def\cI{{\cal I}}
\def\cJ{{\mathscr J}} \def\cK{{\cal K}} \def\cL{{\cal L}}
\def\cM{{\cal M}} \def\cN{{\cal N}} \def\cO{{\cal O}}
\def\cP{{\cal P}} \def\cQ{{\cal Q}} \def\cR{{\cal R}}
\def\cS{{\cal S}} \def\cT{{\cal T}} \def\cU{{\cal U}}
\def\cV{{\cal V}} \def\cW{{\cal W}} \def\cX{{\cal X}}
\def\cY{{\cal Y}} \def\cZ{{\cal Z}}

\def\ua{\underline{\alpha}}
\def\uc{\underline{\phantom{\alpha}}\!\!\!\gamma}
\def\um{\underline{\mu}}
\def\ud{\underline\delta}
\def\ue{\underline\epsilon}
\def\una{\underline a}\def\unA{\underline A}
\def\unb{\underline b}\def\unB{\underline B}
\def\unc{\underline c}\def\unC{\underline C}
\def\und{\underline d}\def\unD{\underline D}
\def\une{\underline e}\def\unE{\underline E}
\def\unf{\underline{\phantom{e}}\!\!\!\! f}\def\unF{\underline F}
\def\unm{\underline m}\def\unM{{\underline M}}
\def\unn{\underline n}\def\unN{{\underline N}}
\def\unp{\underline{\phantom{a}}\!\!\! p}\def\unP{\underline P}
\def\unq{\underline{\phantom{a}}\!\!\! q}
\def\unQ{\underline{\phantom{A}}\!\!\!\! Q}
\def\unH{\underline{H}}

\def\As {{A \hspace{-6.4pt} \slash}\;}
\def\bs {{b \hspace{-6.4pt} \slash}\;}
\def\Ds {{D \hspace{-6.4pt} \slash}\;}
\def\Gts {{\Gt \hspace{-6.4pt} \slash}\;}
\def\ds {{\del \hspace{-6.4pt} \slash}\;}
\def\ss {{\s \hspace{-6.4pt} \slash}\;}
\def\ks {{ k \hspace{-6.4pt} \slash}\;}
\def\ps {{p \hspace{-6.4pt} \slash}\;}
\def\xs {{x \hspace{-6.4pt} \slash}\;}
\def\pas {{{p_1} \hspace{-6.4pt} \slash}\;}
\def\pbs {{{p_2} \hspace{-6.4pt} \slash}\;}
\def\cFs {{{\cal F} \hspace{-6.4pt} \slash}\;}
\def\Dss {{D \hspace{-7.5pt} \slash}\;}
\def\dss {{\del \hspace{-7.0pt} \slash}\;}

\def\Ah{{\hat{A}}}
\def\Dh{{\hat{D}}}
\def\Gh{{\hat{G}}}
\def\Fh{{\hat{F}}}
\def\Ih{{\hat{I}}}
\def\Jh{{\hat{J}}}
\def\Kh{{\hat{K}}}
\def\Lh{{\hat{L}}}
\def\Ph{{\hat{P}}}
\def\Rh{{\hat{R}}}
\def\Vh{{\hat{V}}}
\def\Xh{{\hat{X}}}

\def\ah{{\hat{\a}}}
\def\bh{{\hat{\b}}}
\def\gh{{\hat{\g}}}
\def\dh{{\hat{\d}}}
\def\rh{{\hat{\r}}}
\def\hh{\hat{h}}
\def\uh{\hat{u}}
\def\xh{\hat{x}}
\def\yh{\hat{y}}
\def\ph{\hat{p}}
\def\xih{\hat{\xi}}
\def\chih{\hat{\chi}}
\def\Psih{\hat{\Psi}}
\def\phih{\hat{\phi}}

\def\psit{\tilde{\psi}}
\def\Psit{\tilde{\Psi}}
\def\Psibt{\tilde{\bar{Psi}}}

\def\lambdat{\tilde {\lambda}}
\def\st{\tilde{\sigma}}

\def\delt{\tilde{\delta}}
\def\Phit{\tilde{\Phi}}
\def\Phitb{\overline{\tilde{Phi}}}
\def\tht{\tilde{\th}}
\def\lt{\tilde{\l}}
\def\chit{\tilde{\chi}}
\def\phit{\tilde{\phi}}

\def\At{\tilde{A}}
\def\Bt{\tilde{B}}
\def\Ct{\tilde{C}}
\def\Dt{\tilde{D}}
\def\Et{\tilde{E}}
\def\Ft{\tilde{F}}
\def\Gt{\tilde{G}}
\def\Ht{\tilde{H}}
\def\It{\tilde{I}}
\def\Jt{\tilde{J}}
\def\Qt{\tilde{Q}}
\def\Rt{\tilde{R}}
\def\Mt{\tilde{M }}
\def\Nt{\tilde{N}}
\def\St{\tilde{S}}
\def\Vt{\tilde{V}}
\def\Xt{\tilde{X}}
\def\at{\tilde{a}}
\def\ct{\tilde{c}}
\def\dt{\tilde{d}}
\def\htt{\tilde{h}}
\def\ft{\tilde{f}}
\def\gt{\tilde{g}}
\def\pt{\tilde{p}}
\def\qt{\tilde{q}}
\def\vt{\tilde{v}}
\def\nt{\tilde{n}}
\def\ut{\tilde{u}}
\def\wt{\tilde{w}}
\def\zt{\tilde{z}}
\def\xt{\tilde{x}}
\def\yt{\tilde{y}}
\def\Psit{\tilde{\Psi}}
\def\vphit{\tilde{\varphi}}
\def\tD{\tilde{\D}}

\def\eb{\bar{\epsilon}}
\def\delb{\bar{\partial}}
\def\thb{\bar{\theta}}
\def\mub{\bar{\mu}}
\def\lamb{\bar{\l}}
\def\psib{\bar{\psi}}
\def\sb{\bar{\sigma}}
\def\xib{\bar{\xi}}
\def\chib{\bar{\chi}}

\def\Psib{\bar{\Psi}}
\def\Phib{\bar{\Phi}}
\def\Lamb{\bar{\Lambda}}
\def\Sb{{\overline \Sigma}}
\def\cb{\bar{c}}
\def\hb{\bar{h}}
\def\qb{\bar{q}}
\def\wb{\bar{w}}
\def\ub{\bar{u}}
\def\zb{{\bar{z}}}
\def\Hb{\bar{H}}
\def\Qb{{\bar Q}}
\def\Omegab{\overline{\Omega}}
\def\ob{\overline{\omega}}

\def\Ab{{\overline A}} \def\Bb{{\overline B}} \def\Cb{{\overline C}}
\def\Db{{\overline D}} \def\Eb{{\overline E}} \def\Fb{{\overline F}}
\def\Gb{{\overline G}}
\def\Ib{{\overline I}}
\def\Jb{{\overline J}} \def\Kb{{\overline K}} \def\Lb{{\overline L}}
\def\Mb{{\overline M}} \def\Nb{{\overline N}} \def\Ob{{\overline O}}
\def\Pb{{\overline P}}  \def\Rb{{\overline R}}
 \def\Tb{{\overline T}} \def\Ub{{\overline U}}
\def\Vb{{\overline V}} \def\Wb{{\overline W}} \def\Xb{{\overline X}}
\def\Yb{{\overline Y}} \def\Zb{{\overline Z}}

\def\fb{{\overline f}}
\def\gb{{\overline g}}
\def\mb{{\overline m}}
\def\lb{{\overline l}}
\def\yb{{\overline y}}

\def\ldel{{\overleftarrow{\del}}}
\def\rdel{{\overrightarrow{\del}}}
\def\ldeldel{{\overleftarrow{\del^2}}}
\def\rdeldel{{\overrightarrow{\del^2}}}
\def\ldelb{{\overleftarrow{\bar{\del}}}}
\def\rdelb{{\overrightarrow{\bar{\del}}}}

\def\ba{{\bf a}}
\def\bk{{\bf k}}
\def\bl{{\bf l}}
\def\bp{{\bf p}}
\def\bq{{\bf q}}
\def\br{{\bf r}}
\def\bt{{\bf t}}
\def\bu{{\bf u}}
\def\bv{{\bf v}}
\def\bx{{\bf x}}
\def\by{{\bf y}}
\def\bA{{\bf A}}
\def\bR{{\bf R}}
\def\bV{{\bf V}}

\def\bz{{\boldsymbol{\zeta}}}

\def\bone{{\bf 1}}

\def\va{{\vec a}}
\def\vk{{\vec k}}
\def\vp{{\vec p}}
\def\vq{{\vec q}}
\def\vx{{\vec x}}
\def\vy{{\vec y}}
\def\vu{{\vec u}}
\def\vv{{\vec v}}
\def \vH{{\vec H}}
\def \vg{{\vec g}}

\def\vs{{\vec \sigma}}
\def\vtau{{\vec \tau}}

\newcommand{\ov}[1]{\overrightarrow{#1}}

\def\frA{\mathfrak{A}}
\def\frB{\mathfrak{B}}
\def\frC{\mathfrak{C}}
\def\frD{\mathfrak{D}}
\def\frE{\mathfrak{E}}
\def\frF{\mathfrak{F}}
\def\frG{\mathfrak{G}}
\def\frH{\mathfrak{H}}
\def\frM{\mathfrak{M}}
\def\frN{\mathfrak{N}}
\def\frR{\mathfrak{R}}
\def\frW{\mathfrak{W}}

\def\fra{\mathfrak{a}}
\def\frb{\mathfrak{b}}
\def\frf{\mathfrak{f}}
\def\frg{\mathfrak{g}}
\def\frh{\mathfrak{h}}
\def\frl{\mathfrak{l}}
\def\frs{\mathfrak{s}}
\def\fri{\mathfrak{i}}
\def\frj{\mathfrak{j}}

\def\ma{\mathfrak{a}}
\def\mg{\mathfrak{g}}
\def\mh{\mathfrak{h}}
\def\mR{\mathfrak{R}}
\def\mN{\mathfrak{N}}

\newcommand{\nn}{{\nonumber}}

\def\d{\delta}\def\D{\Delta}\def\ddt{\dot\delta}

\def\pa{\partial} \def\del{\partial}
\def\xx{\times}
\def\uno{\mbox{1 \kern-.59em {\rm l}}}

\def\trp{^{\top}}
\def\inv{^{-1}}
\def\dag{\dagger}
\def\pr{^{\prime}}

\def\rar{\rightarrow}
\def\lar{\leftarrow}
\def\lrar{\leftrightarrow}

\newcommand{\0}{\,\!}      
\def\one{1\!\!1\,\,}
\def\im{\imath}
\def\jm{\jmath}

\newcommand{\tr}{\mbox{tr}}
\newcommand{\slsh}[1]{/ \!\!\!\! #1}

\newcommand{\1}{\mbox{1}\hspace{-0.25em}\mbox{l}}

\def\vac{|0\rangle}
\def\lvac{\langle 0|}

\def\hlf{\frac{1}{2}}
\def\ove#1{\frac{1}{#1}}
\newcommand{\hot}[1]{\frac{#1}{2}}

\def\Box{\square}
\def\CC {\mathbb{C}}
\def\FF {\mathbb{F}}
\def\RR{\mathbb{R}}
\def\NN{\mathbb{N}}
\def\ZZ{\mathbb{Z}}
\def\bb#1{{\bf #1}}
\def\bcomment#1{}
\def\bfhat#1{{\bf \hat{#1}}}
\def\VEV#1{\left\langle #1\right\rangle}

\newcommand{\ex}[1]{{\rm e}^{#1}} \def\ii{{\rm i}}

\newcommand{\lrbrk}[1]{\left(#1\right)}
\newcommand{\lrsbrk}[1]{\left[#1\right]}
\newcommand{\sfrac}[2]{{\textstyle\frac{#1}{#2}}}

\def\stw{{\sqrt{2}}}

\def\rf {{\rm f}}
\def\ri {{\rm i}}
\def\rj {{\rm j}}
\def\rn {{\rm n}}
\def\rk {{\rm k}}
\def\rl {{\rm l}}
\def\rr {{\rm r}}
\def\rs {{\scriptscriptstyle \rm S}}
\def\rt {{\scriptscriptstyle \rm T}}

\def\rQ {{\scriptscriptstyle \rm \cQ}}
\def\rR {{\scriptscriptstyle \rm \cR}}

\def\cQb{{\cal \Qb}}
\def\cRb{{\cal \Rb}}
\def\cWb{{\cal \Wb}}

\def\fd {{\rm N}}
\def\afd {{\overline{\rm N}}}

\def \II {I\hspace{-.1em}I\hspace{.1em}}
\def \IIA {\mbox{\II A\hspace{.2em}}}
\def \IIB {\mbox{\II B\hspace{.2em}}}
\def \gs {g^s}
\def \ls {\lambda^s}

\def \I {{\cal I}}
\def \qs {q\hspace{-.53em}/\hspace{.15em}}
\def \ks {k\hspace{-.53em}/\hspace{.15em}}
\def \YM {{\mbox{\tiny YM}}}
\def \gym {g_{\YM}}

\def \Lc {\L_c}
\def\IR{\relax{\rm I\kern-.18em R}}
\def \id {{\bf 1}}

\def\cci{\ell}
\def\ccj{\ell'}

\def\bbq{\pmb{q}}
\def\bom{\pmb{\o}}
\def\bJ{\pmb{J}}
\def\bM{\pmb{M}}
\def\bB{\pmb{B}}
\def\bn{\pmb{n}}
\def\bE{\pmb{E}}

\newcommand{\rrr}[1]{\vskip 0.2cm \noindent{\it #1} ---}

\begin{document}
\hfill{ NCTS-TH/1802}
\title{ Weyl Anomaly Induced Current in Boundary Quantum Field Theories}
\author{Chong-Sun Chu${}^{1,2}$}
\author{Rong-Xin Miao${}^{1,3}$ }
\affiliation{${}^1$ National Center for Theoretical Sciences, National Tsing-Hua
  University, Hsinchu 30013, Taiwan\\
${}^2$ Department of Physics, National Tsing-Hua
 University, Hsinchu 30013, Taiwan\\
 ${}^3$ School of Physics and Astronomy, Sun Yat-Sen University, Zhuhai, 519082, China}


\begin{abstract}
  We show that when an external
  magnetic field
  parallel to the boundary
  is applied, Weyl anomaly give rises to a new
  anomalous current
  in the vicinity of the boundary.
  The induced current is a magnetization current in origin:
  the movement of the virtual charges near the boundary give rise to
  a non-uniform magnetization of the vacuum and hence a magnetization current. 
  Unlike other previous studied anomalous
  current phenomena such as the chiral magnetic effect or the chiral vortical
  effect, this induced current
  does not rely on the presence of a
  material system and can occur in vacuum.  
  Similar to the Casimir effect, 
  our discovered phenomena
  arises from the effect of the boundary on the
  quantum fluctuations of the vacuum.
  However this induced current is
pure quantum mechanical and
has no  classical limit.
We briefly
comment on how this
  induced
  current  may be observed experimentally.

\end{abstract}

\maketitle


\rrr{Introduction}
Quantum anomaly induced current is an
interesting
phenomena. Much has been discussed in
the literature \cite{review}. A number of such effects are known. The
famous one is 
the chiral magnetic effect (CME) \cite{Vilenkin:1995um,
  Vilenkin:1980fu, Giovannini:1997eg, alekseev, Fukushima:2012vr} which refers
to the generation of currents  parallel to an external
magnetic field $\bB$.
The
chiral vortical effect (CVE) \cite{Kharzeev:2007tn,Erdmenger:2008rm,
  Banerjee:2008th,Son:2009tf}
refers to the generation of a current
due to rotational motion in the charged fluid.  
The induced currents take the form
\be \label{j1}
\bJ_V = \s_{(\cB) V} \bB + \s_{(\cV) V} \bom, \quad
\bJ_A = \s_{(\cB)  A} \bB+ \s_{(\cV) A} \bom,
\ee
where 
$\s_{(\cB) V} = \frac{e \mu_A}{2 \pi^2}$,
$\s_{(\cB) A} = \frac{e \mu_V}{2 \pi^2}$
are the chiral magnetic conductivities,
$\s_{(\cV) V} = \frac{ \mu_V \mu _A }{\pi^2}$,
$\s_{(\cV) A} = \frac{\mu_V^2+\mu_A^2}{2\pi^2} + \frac{T^2}{6}$
are the chiral vortical conductivities, 
$\mu_A, \mu_V$ are the chemical potentials  and $T$ is the temperature
 of the medium. The chemical potential dependent induced current
 arises as a result of an imbalance in the left and right moving
modes due to the axial anomaly, while the temperature dependent part
comes from the gravitational anomaly \cite{Landsteiner:2011cp}. 
More recently, it has
also been pointed out that anomalous
current also occurs in a conformally flat gravitational spacetime
due to Weyl
anomaly \cite{Chernodub:2016lbo, Chernodub:2017jcp}.
It should be noted that these anomalous
current occurs only in a material system where the chemical potentials
are non-vanishing, or in a curved spacetime. Since axial anomaly is an
intrinsic property of Quantum Field Theory (QFT) which is present
even in flat
spacetime and in  vacuum,
it is natural to ask whether
the phenomena of anomalous current may also occur in
flat spacetime due to quantum fluctuation of the vacuum.

The Casimir effect is one of the most well known manifestation of the
quantum fluctuation of electromagnetic
vacuum in the presence of boundary
\cite{Casimir:1948dh,Plunien:1986ca,Bordag:2001qi}.
Recently the Casimir effects has been
analyzed in full generality for arbitrary shape of boundary and
for arbitrary spacetime metric, and
new universal relations between the Casimir coefficients
and the boundary central charge in a boundary conformal
field theory have been discovered \cite{Miao:2017aba}.
The presence of boundary has also many other interesting physical
consequences, e.g. renormalization
group flows and critical phenomena \cite{Cardy:2004hm} or
the topological
insulator \cite{Hasan:2010xy} etc.

In this paper, we show that for
a general class of boundary quantum field theory
(BQFT) with $U(1)$ gauge symmetry,
the
quantum Weyl anomaly of the theory induces a  new kind of
induced current
near the boundary.
Consider a general BQFT defined on a four dimensional spacetime
manifold $M$ with coordinates $x^\m$, and has boundary $\del M$ with
coordinates $y^a$.
The Weyl anomaly can be defined as the difference
between the trace of renormalized stress tensor and the renormalized
trace of stress tensor \cite{Duff:1993wm,Brown:1976wc}.
We find it useful to introduce the following
integrated Weyl anomaly
\begin{eqnarray}
  \label{A0}
\mathcal{A}=\int_M \sqrt{g} \Big[ g^{\m\n} \la T_{\m\n}\ra-\la g^{\m\n}
  T_{\m\n}\ra\Big].
\end{eqnarray}
$\cal A$ is
equal to the variation of the effective action with respect to 
constant re-scaling of the metric\cite{Deser:1993yx}. 
For simplicity, we focus on QFT which are covariant, gauge invariant,
unitary and renormalizable, e.g. QED.
By ``renormalizable'', we mean, in the
sense of perturbation theory, that all the coupling constants are
of non-negative
mass dimension. We also assume that the
Weyl anomaly depends on only the positive powers of the coupling
constants (including the mass $m$), so that it has a well-defined
limit when we turn off the coupling constants.  For this class of QFT,
$\cA$ takes the following form \cite{Duff:1993wm,note}
\be
 \label{A1}
\mathcal{A}=\int_M \sqrt{g} [ b_1 F_{\m\n}F^{\m\n}+O(R^2)]+\int_{\partial
  M} \sqrt{h} O(Rk).
\ee
Here $O(R^2)$ denotes terms constructed out of the bulk curvature
tensor, including terms with positive powers of coupling constants;
e.g. $R^2, R_{\m\n}R^{\m\n}, R_{\m\n\a\b}R^{\m\n\a\b}, \Box R, m^2 R, m^4, \cdots$,
and  $O(Rk)$ denotes the
boundary Weyl anomaly \cite{Fursaev:2015wpa,Herzog:2015ioa}
that is constructed out of the boundary 
curvature tensor and the exterior curvature of the boundary.
$b_1$ is the bulk central charge which govern
the gauge field contribution to the Weyl anomaly \eq{A1}.
For the normalization of the gauge field kinetic term
$ S = -1/(4e^2) \int F^2$, $b_1$ is related to the beta function as 
$b_1 = -\frac{\b(e)}{2 e^3}$ \cite{Peskin}.
Below we show that for
general BQFT as specified above,
the expectation value of the induced current
at a distance $x$ very close to the boundary \cite{note1}
is given by
\be
  \label{current1}
  \la \bJ \ra = \frac{e^2 c}{\hbar}
  \frac{4 b_1 \bn \times \bB}{x} , \quad x \sim 0,
  \ee
  where $\bn$ is the inner normal to the boundary.
  The current \eq{current1} is a magnetization current
  $\bJ = \nabla \times \bM$ and corresponds to a quantum magnetization
  \be \label{M}
\la \bM \ra =  \frac{e^2 c}{\hbar} 4 b_1 \log x \; \bB
  \ee
  of the vacuum.
  It is remarkable that the
anomalous current \eq{current1}
and the vacuum magnetization \eq{M}
takes place even in flat
spacetime and at zero temperature.
These are pure quantum effect since it
is inversely
proportional the Planck constant
and has no classical limit $\hbar \to 0$. 
The induced current is measured by quantum Hall conductance 
$\s_H = e^2 / \hbar$  which
govern the quantum Hall effect. In fact the current
(\ref{current1})
is in
resemblance to the quantum Hall effect except that the
current now is
parallel to the boundary instead of perpendicular to the boundary as in the
case of the standard Hall effect.
One may therefore refer to 
(\ref{current1})
as an {\it Anomalous Quantum Hall Effect} \cite{note2}.

\rrr{Physical Picture}
To understand the physical origin of the current \eq{current1}
and the magnetization \eq{M}. 
Let us consider for simplicity
the set up of a  BQFT in flat spacetime with a flat boundary.
Consider a point $P$ at distance $x$ from the boundary. We are
interested in the amount of charges passing through $P$ due to 
vacuum process of virtual particle
creation and annihilation.
Suppose there is a magnetic field
normal to (pointing out of) the figure,
  the charged particles will move along circles
  due to the Lorentz force.
  If there is no boundary, the virtual particle pairs created by
  quantum fluctuations at $O'$ would annihilate at $P$ after moving
  along the dotted circle. This give rises to a
  transport of charges to the right.
  This is however precisely  canceled by the movement of charges due to quantum
  fluctuation at the point $O''$. Summing over all possible locations of
  the source points,
  it is clear that there is no net transport of charges induced at
  $P$.
  The situation is different when
  there is a boundary. In this case, those contribution from  source
  points at $x<0$ are missing.  This leads to a net amount of charges
  moving to $-y$ direction.
  In addition, vacuum pairs created at source point $O'''$ could now
  reach $P$ due
  to
  (virtual) reflection of the boundary. What exactly happens,
  perfect reflection or partial absorption,  will
  depend on the boundary condition. But in any case there will be a net
  separation
  of charges and this contributes a transport of charges to the
  $+y$ direction.

The current \eq{current1} can also be understood as a result of the
magnetic response of the vacuum
to the presence of boundary. As we noted already,
quantum fluctuation of the vacuum leads to
temporary creation of virtual pair of charged particles, which are then
guided to move on circles in the presence of a magnetic field. As a
result, tiny current loops are formed with the positively and negatively charged
virtual particles 
contributing in the same way to the magnetic dipole moment.
Summing all these contribution results in  a total
magnetization $\bM$ of the vacuum.
When there is no boundary, $\bM$ is just an infinite
constant that
can be subtracted away by renormalization and
the renormalized vacuum magnetization $\la \bM \ra =0$
has no physical effect. When there is a
boundary, it is clear that the renormalized $\la \bM \ra$ is zero far away
from the boundary, but become nontrivial near the boundary. This is very much
like the Casimir effect. 
The magnetization \eq{M}
of the vacuum is a new effect and occurs only because of
the presence of the boundary.
Let us now turn to
the rigorous QFT derivation.

\begin{figure}[t]
\centering
\includegraphics[width=7cm]{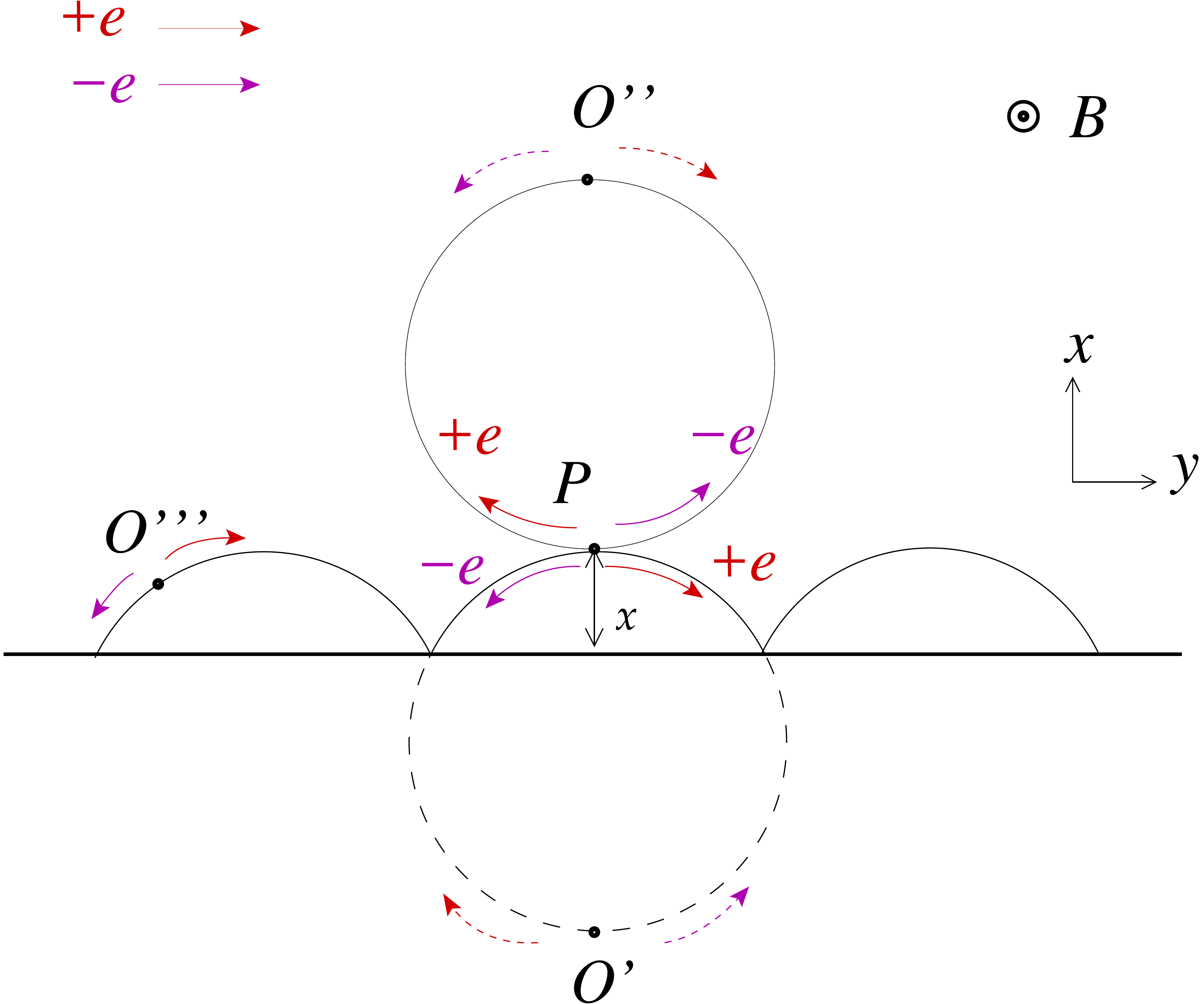}
\caption{Induced current from virtual pair creation in presence of
  boundary. }
\end{figure}


\rrr{Rigorous Derivation} 
We start with a proper
analysis of the structure of the renormalized
current $J^\m$ near the boundary. In general, for a BQFT,
the renormalized current is generally singular near
the boundary and the expectation value takes the asymptotic form
near $ x \sim 0$:
\begin{eqnarray}\label{current0}
  \la J_\m\ra = \frac{1}{x^3}  J^{(3)}_{\m}+\frac{1}{x^2}  J^{(2)}_{\m}
  +\frac{1}{x}  J^{(1)}_{\m}
  + \log x J^{(0)} + \cdots,\;\; 
\end{eqnarray}
where
$\cdots$ denotes
terms regular at $x=0$, and $J^{(n)}_{\m}$
depend only the background geometry, the background vector field strength
and the type of fields under consideration.
Hereafter we will drop the symbol $\la \; \ra$ for the expectation value. 
A 
similar expansion has been considered for the renormalized stress tensor
\cite{Deutsch:1978sc}.
We consider current that is conserved
($ D_\m  J^\m  = 0$)
up to  possibly an anomaly term. Since this
term
is finite,
it is irrelevant to the divergent part of renormalized current
(\ref{current0}).
As a result, we obtain
the gauge invariant solution
\begin{equation}\label{solnJ1}
\begin{split}
  &J^{(3)}_{ \m}=0, \quad \ J^{(2)}_{\m}=0,\\ &J^{(1)}_{\m}=
  \a_1  F_{\m\n} n^\n
  +\a_2 \mathcal{D}_\m k+\a_3 \mathcal{D}_\n k^\n_\m
 +\a_4  \star F_{\m\n}\, n^\n
\end{split}
\end{equation}
where
$F_{\m\n}$, $\star F_{\m\n}$, $n_\m$, $\mathcal{D}_m$, $k_{\m\n}$ and
$h_{\m\n}$ are respectively the background field strength, 
Hodge dual of field strength, 
the normal vector, induced
covariant derivative, extrinsic curvature and induced metric of the
boundary. 
Note that  in \eq{solnJ1} we have re-expressed
$n^\m R_{\m\n}h^{\m}_\n$ in terms of extrinsic
curvatures by using the Gauss-Codazzi equation
$n^\m R_{\m\n}h^{\n}_\g =\mathcal{D}_\m k^\m_\g- \mathcal{D}_\g k$.
Here the coefficients $\a_i$ are arbitrary and
the expression \eq{solnJ1} gives the most
general form of boundary behavior of the
current that is consistent with the conservation law and gauge invariance,
We will now show that
these current coefficients are indeed completely fixed 
by the central charges of the theory. 

To establish this result, let us follow an
observation of \cite{Miao:2017aba} which allows one to relate the variation of
$\cal A$ with the asymptotic form of the stress tensor
near the boundary.
For the present case of current, we have the relation
\be \label{key}
(\d \cA )_{\del M_\e} = \Big(\int_M \sqrt{g} J^\mu \d A_\mu
\Big)_{\log \frac{1}{\e}},
\ee
where a regulator $x \geq \e$ to the boundary is introduced for the integral
on the
right hand side (RHS) of \eq{key}. The relation \eq{key}
identifies the boundary contribution of
the variation of the
integrated anomaly $\cal A$
under an arbitrary variation of the gauge field $\d A_\m$ with  the UV
logarithmic divergent part of the integral
involving the expectation value
$J^\m$ of the renormalized $U(1)$ current.
The power of the relation \eq{key} lies in the fact that the left hand
side of \eq{key} is a total variation
and impose
constraints on the RHS of \eq{key} that are
powerful enough to
to fix completely the asymptotic behavior
of the current in terms of the Weyl anomaly of the theory.
We refer the readers to the appendix for the derivation of
this key relation (\ref{key}).

Now let
us use \eq{key} to fix the current coefficients.
To proceed, let us consider the metric written in the Gauss
normal coordinates $ds^2=dx^2+ \left(h_{ab}-2x k_{ab}+ x^2
q_{ab}+ \cdots \right)dy^a dy^b$,
where $x\in [0,+\infty)$ and
  $n_\m=(1,0,0,0)$ is the inward pointing
  normal vector. We also choose a gauge $A_x=0$ and
  expand the gauge field about the boundary: 
  $A_b=a_b+ xA^{(1)}_b+ \cdots$. 
  Taking the variation of Weyl anomaly (\ref{A1}) with
  respect to the gauge field,
  we have
$ ( \d\cA)_{\del M}  =   4b_1 \int_{\partial M}\sqrt{h}   F^b{}_n \, \d a_b$.
  Next, we substitute
\eq{current0}, \eq{solnJ1}
into the RHS of (\ref{key}),
integrate over $x$
and select the logarithmic divergent term, we obtain
$ \Big(\int_M \sqrt{g} J^\mu \d A_\mu
\Big)_{\log 1/\e} = \int_{\partial  M}   \sqrt{h}
   (\a_1 F^b{}_{n}+\a_2 \mathcal{D}^b k+\a_3 \mathcal{D}_j k^{jb}
 + \a_4 \star F^b{}_n )\delta a_b$.
As a result, we obtain, for unitary QFT without the parity
odd anomaly term \cite{note},
$\a_1=4b_1, \a_2 = \a_3 = \a_4 = 0$, 
and 
our main result for the expectation value of the current 
near the boundary:
\begin{eqnarray}
  \label{current2}
 J_b  = \frac{4 b_1 F_{b n}}{x}, \quad x \sim 0,
\end{eqnarray}
We emphasis that the current \eq{current1}
does not involve any on-shell charged particle as
we were considering the vacuum state and there is no Schwinger effect
for magnetic field. Instead  the 
induced current should be identified with a magnetization
current as a result of the magnetization \eq{M} of the vacuum. This can be derived
directly without first referring to the current \eq{current1}
by using the magnetic coupling 
$ S = \int_M \sqrt{g} \bM \cdot \bB$
and the relation
\be \label{AJ}
(\d \cA)_{\del M_\e}=
\left(\int_M \sqrt{g} \bM\cdot \d \bB \right)_{\log \frac{1}{\e}}.
\ee
By considering a variation
$\d B_z = \delta(x -\e) \d f(y,z)$ that is localized on the boundary 
$\del M$,
one obtain \eq{M}.

The universal laws \eq{current1} and \eq{M}
hold for general BQFTs which are covariant, gauge invariant, unitary
and renormalizable, or equivalently, for BQFTs whose Weyl anomaly
is given by (\ref{A1}).  
Several comments are in order.
{\tt 1}. Since \eq{current1} and \eq{M}
depend on only the bulk central charge instead of
boundary central charge, it is independent of the choices of boundary
conditions. Thus the current is more universal than the renormalized
stress tensor near the boundary which depends on boundary conditions
\cite{Deutsch:1978sc,Kennedy:1979ar,Kennedy:1981yi,Miao:2017aba}.
{\tt 2}.
     The magnitude  of the induced current is much larger than that of the
stress tensor. To see this, let us recover the
units in the formula. We have
\begin{eqnarray}
  \label{currentstress}
J_b  = \frac{e^2 c}{\hbar} \frac{4 b_1 F_{b n}}{x},\ \ \ 
T_{ab} = \hbar c \frac{d_1 h_{ab}}{x^4},
\end{eqnarray}
where $e$ is the charge, $c$ is the speed of light, $\hbar$ is the
Planck constant, 
$b_1, d_1$ are dimensionless constants and
$h_{ab}$ is the boundary metric. We have re-scaled
$F_{\m\n}\to e F_{\m\n}$ so that the field strength is related to
electric field and magnetic field in the usual manner:
$E_i=c F_{i0}, B_i= \frac{1}{2}\epsilon_{ijk}F^{jk}$.
{\tt 3}.
Our result shows that 
constant magnetic field parallel to the boundary  can induce a
current \eq{current1}.
As we illustrated
above, the boundary
is crucial in realizing a separation
of charges which result in the induced anomalous current
and in the non-uniform magnetization for the vacuum.
{\tt 4}.
We emphasis that our current is not due to on-shell movement of charges, but
transport of virtual charges as a result of non-uniform vacuum magnetization.
As such our current does not obey Ohm's law and is not dissipative. It
does not require an energy source to support it.
{\tt 5}. The result \eq{current1} is for a single boundary. For a real system with
finite extent, e.g. a rectangular slab with two parallel boundaries,
we will have current of the same form near each 
boundary components of the system.
The total current is zero and satisfy the Bloch theorem \cite{Yamamoto:2015fxa}. 
{\tt 6}. The relation
\eq{current2}
also implies an induced charge density
$\rho = \frac{e^2}{\hbar} \frac{4b_1 E}{x}$
near the boundary.
Here $\bE = E \pmb{e}_x$.
{\tt 7.} Our results \eq{current1} and \eq{M}
were derived for the vacuum. In a material system, one need to take into account of
the presence of charge carriers and non-vanishing conductivity of the media.
The direct field theory analysis seems rather
complicated. However due to the close relation with the Weyl anomaly, we expect
that these results will continue to hold.
In \cite{ChuMiao}
we use a holographic model
to study the effect of conductivity, and we
find that the current and the magnetization
are not modified in the leading order of closeness to the boundary.

\rrr{Story of Free QFT}
Our general result \eq{current2} is  verified by free BQFT. For
simplicity, let us consider complex
scalar field with the action
$ I=-\int_M \sqrt{g}[ (D^\m \phi)^* D_\m \phi+ E \phi^*\phi ]$,
where $D_\m= \del_\m+i A_\m$ are gauge invariant covariant derivatives
and $E$ are functions including only the coupling constants with
non-negative mass dimension. For example, we can have $E=m^2+ \lambda_0
R+...$. However we exclude the terms like
$E=\lambda_{1}F_{ij}F^{ij}+\lambda_{2}R^2$ since they are
non-renormalizable. In general, there are two kinds of boundary
conditions for the scalar \cite{McAvity:1990we}:
Dirichlet BC ($\phi|_{\partial M}=0$) and
the Robin BC ( ($D_n + \psi)\phi|_{\partial M}=0$).
Here the function $\psi$ defines a renormalizable theory, for
example, $\psi=2 \lambda_0 k+m f(y)+ \cdots $.
For a free complex scalar field theory,
the expectation value of the current near the boundary has been derived in
\cite{McAvity:1990we} using heat kernel expansion. The result is
$J_b =-\frac{F_{b n}}{24\pi^2x}$
for both Dirichlet BC and Robin BC. The Weyl anomaly for
the complex scalar theory
can be derived as the heat-kernel coefficient
$a_4$ \cite{Vassilevich:2003xt,Duff}. In this way, we get the Weyl
anomaly (\ref{A1}) with the central charge
$b_1= -1/96\pi^2$.
It is clear that the
obtained current
indeed satisfies our derived universal law (\ref{current2}).
From this simple example, we have learned two important
facts. First, the near-boundary current is indeed independent of the
choices of boundary conditions. 
Second, the universal law
(\ref{current2}) works for not just BCFT,  but also for
more general QFT. The only constraints
we impose on the functions $E, \psi$  are that they define a
renormalizable theory. In particular,
the theory need not be conformal invariant with $E=\frac{1}{6}R$,
$\psi=\frac{1}{3}k$.

\rrr{Finite Total Current}
Similar to the case of stress tensor
\cite{Kennedy:1979ar,Dowker:1978md, Miao:2017aba}, there are boundary
contributions to the current which make  the total
current finite. To see this, 
consider the gauge variation of finite part of the
effective action. Due to gauge invariance, we obtain the conservation laws
$D_\m J^\m=0$ in the bulk and
$\cD_a j^a= - J^n$ on the boundary.
From the bulk current conservation and (\ref{current2}),
we get
$J_n=4b_1 \cD_b F^{b}_{\ n} \ln x+O(1)$. Substituting $J_n$ into 
the boundary conservation law, we obtain the boundary
current
$j_b=4b_1 F_{b n} \ln \epsilon.$
As a result, we have 
\be\label{currenttotal}
J_b =\frac{4 b_1 F_{b n}}{x}+\delta(x;\partial M)  4 b_1 F_{b n}\ln\epsilon+O(1).
\ee
where we have shifted the boundary from $x=0$ to a position $x=\e$.
It is remarkable that the
boundary current obtained from the
conservation law 
automatically
yields the total current \eq{currenttotal}
which 
represent a finite flow of charge through any interval
in the normal direction.

\rrr{On Experimental Observation}
Our current \eq{current1} can be observed by measuring the
magnetic response of the vacuum to external field in the
presence of boundary.
We have shown that the renormalized current
and  the quantum magnetization are independent of the
choices of well-defined boundary conditions (BC). By `well-defined BC', we means no current can flow out the boundary.
 The insensitivity of boundary
conditions would decrease the difficulty in experiments. 
In reality since modes with  sufficiently high
frequencies would penetrate the 
boundary,
this corresponds to 
an effective length cutoff and our formula (\ref{current2}) will work
well only for
$x>\epsilon$ with the cut off naturally being the lattice length
$a_{\rm lattice}$ of the material in consideration.
Consider, for simplicity, a constant magnetic field $B$ and constant
temperature $T$ for the material. On the other hand,
the formula (\ref{current2}) applies 
only to the region close enough to the boundary such that $x<
x_{\text{max}}=\text{min}\left(\hbar c/(k T), \hbar/(c
\ m_{\text{eff}}), \sqrt{\hbar/(e B)}\right)$,
where $m_{\text{eff}}$ is the effective
mass of the charged particle. Taking
$T =300 {\rm K}$, $m_{\text{eff}}=m_e$ to be the mass of
electron and $B=0.01{\rm T}$, we have $x_{\text{max}}\sim\text{min}\left(
10^{-5}{\rm m}, 10^{-13} {\rm m}, 10^{-6} {\rm m}\right)$,
which shows that the large
mass of charged particle is the main obstruction to experimental
observation of the phenomena. Thus
one must try to decrease the effective mass in materials in order to
satisfy $\e < x_{\text{max}}$. Fortunately, the
availability of charge carriers with zero effective
mass in graphene \cite{graphene} and Dirac or Weyl semimetals \cite{WeylDirac}
makes these systems
a more promising setup for
experimental observation of this induced current phenomena.

\rrr{Conclusions and Discussions}
In this letter, we 
show that for general four dimensional BQFTs
which are  gauge invariant, unitary and renormalizable,
the renormalized current takes the
universal form (\ref{current2}) near the boundary. 
This covers fundamental theories such as QED, as well as
many typical condensed matter systems of interests.
The induced
current is independent of the boundary conditions
and the states of BQFT, and depends
only on the beta function of the theory.
Since the  current is proportional to 
the quantum Hall conductance $e^2/\hbar$, it is
potentially large enough to be measured experimentally.
It is interesting
to perform experiment to observe this effect. 
It is also interesting to look for suitable implication of this effects
for other physical systems such as astronomical objects or branes in
string theory. 
Our discussions can be easily generalized to
system with background non-Abelian
gauge field
and with 
spacetime dimensions other than four.
See the appendix
for the expectation value of current
in dimensions other than four.
We note however that only
in four dimensions
is the
near boundary value of the current determined
universally
by the bulk central charge and is independent of boundary conditions.

 \section*{Acknowledgements}
We thank Bei-Lok Hu, Yan Liu, Jian-Xin Lu, Daw-Wei Wang and Ling-Yan Hung
for useful discussions and comments.
This work is supported in part by 
NCTS and the grant MOST
105-2811-M-007-021 of the Ministry of
Science and Technology of Taiwan. Rong-Xin Miao thank the funding of Sun Yat-Sen University.

\appendix

\section{Supplementary Information}

\subsection{1. The derivation of key formula}
Consider a BQFT with a well defined effective action.
The integrated
Weyl anomaly $\cA$ defined by (\ref{A0})
can be obtained as the logarithmic
UV divergent term of the effective action,
\be \label{I0}
I = \cdots + \cA \log (\frac{1}{\e})  + I_{\rm finite},
\ee
where $\cdots$ denotes terms which are UV divergent in powers of the
UV cutoff $1/\e$, and $ I_{\rm finite}$ is the renormalized, UV finite
part of the
effective action.
To derive this result,
let us consider a constant Weyl transformation
$g_{\mu\nu} \to \exp(2 \omega) g_{\mu\nu}$.
Under this transformation, the UV cutoff
transforms as $\e \to \exp(\omega) \e $ and 
the variation of effective action (\ref{I0}) becomes
\be \label{I01}
\delta_{\omega}I = \cdots + \omega (-\cA   +\int_M \sqrt{g}
\la T^{\mu\nu}\ra g_{\mu\nu})+O(\omega^2),
\ee
where we have used $\delta_{\omega} \cA=0$ and
$\delta_{\omega} I_{\rm finite}= \omega \int_M \sqrt{g}
\la T^{\mu\nu}\ra g_{\mu\nu}+O(\omega^2).$
On the other hand, by definition we have 
\be \label{I02}
\delta_{\omega}I = \frac{1}{2}\int_M \sqrt{g}
\hat{T}^{\mu\nu} \delta_{\omega}g_{\mu\nu}
= \omega \int_M \sqrt{g} \hat{T}^{\mu\nu}g_{\mu\nu}+O(\omega^2),
\ee
where $\hat{T}^{\mu\nu}$
is the non-renormalized stress tensor.
We use the hatted  symbol (e.g. $\hat{T}_{\m\n})$
to denote non-renormalized quantity
and un-hatted symbol (e.g. $T_{\m\n}$) to denote renormalized quantity.
Separating  $\hat{T}^{\mu\nu}g_{\mu\nu}$ into the renormalized UV finite part
$\la \hat{T}^{\mu\nu}g_{\mu\nu} \ra$ and divergent part, we have 
\be \label{I03}
\delta_{\omega}I = \cdots+ \omega \int_M \sqrt{g}
\la \hat{T}^{\mu\nu}g_{\mu\nu} \ra+O(\omega^2).
\ee
Comparing the finite part of (\ref{I01}) and (\ref{I03}), we
obtain  the expression (\ref{A0}) for 
$\cal A$ and hence our claim.

Now we are ready to prove the result \eq{key} quoted in the main text of
this letter. 
Inspired by \cite{Lewkowycz:2014jia,Dong:2016wcf},
let us regulate  the effective action by excluding from its volume
integration a small  strip of  geodesic distance $\epsilon$ from the boundary.
Then there is no explicit boundary
divergences in this form of the effective action, however there are boundary
divergences implicit in the bulk effective action
which is integrated up to distance $\epsilon$.
The variation of effective action with respect to the vector is given by
\begin{eqnarray} \label{key1}
\delta I= \int_{x \ge \epsilon} \sqrt{g} \hat{J}^{\mu}\delta A_{\mu}
\end{eqnarray}
where $\hat{J}^{\mu}=\frac{\delta I}{\sqrt{g}\delta A_{\mu}}$ is the
non-renormalized bulk current.
The renormalized bulk current is defined by the difference
of the  non-renormalized bulk current against a reference one
\cite{Deutsch:1978sc}:
\begin{eqnarray}
  \label{keyJ3}
J^{\mu}=\hat{J}^{\mu}-\hat{J}_0^{\mu},
\end{eqnarray}
where $\hat{J}_0^{\mu}$ is the
non-renormalized current
defined for the same CFT without boundary. It is
\begin{eqnarray} \label{key2}
\delta I_{0}= \int_{x \ge \epsilon} \sqrt{g} \hat{J}_0^{\mu}\delta A_{\mu},
\end{eqnarray}
where $I_0$ is the effective action of  the  CFT 
with the boundary removed, hence the integration
over the region $x\ge \epsilon$.
Subtract (\ref{key2}) from (\ref{key1}) and focus on only the
logarithmically divergent terms, we obtain our key formula 
\begin{eqnarray} \label{key-1}
  (\delta \mathcal{A})_{\partial M} = \left(
  \int_{x \ge \epsilon} \sqrt{g}J^{\mu}\delta
A_{\mu} \right)_{\log (1/\epsilon)},
\end{eqnarray}
where $(\delta \mathcal{A})_{\partial M}$ is the boundary terms in the variations
of Weyl anomaly and $J^{\mu}$ is the renormalized bulk current.
In the above derivations, we have used the fact that $\delta I$
and $\delta I_0$ have
the same bulk variation of Weyl anomaly so that 
\begin{eqnarray} \label{key3}
(\delta \mathcal{A})_{\partial M}= (\delta I-\d I_0)_{\log (1/\epsilon)}.
\end{eqnarray}
We remark that we have considered global Weyl rescaling here and this is
sufficient for the derivation of our results. For general local Weyl 
rescaling, the transformation properties of the effective action in the
presence of boundaries can be analyzed in the line of \cite{Dowker:1989ue}
and we will leave it for future work.

\subsection{2. Renormalized currents in $d \neq 4$}

Consider  BQFT in $d$-dimensional spacetime,
flat for simplicity.
In higher dimensions, it is expected that the renormalized current
takes the following form
\begin{eqnarray}
  \label{currenthigherd}
  \la J_\m\ra = \frac{\alpha_1 F_{\m n}}{x^{d-3}}
  + \a_2 \frac{\del^\n F_{\m\n}}{x^{d-4} }+ \cdots , \quad  x \sim 0,
\end{eqnarray}
near the boundary.
We claim that for $d>4$, $\alpha_1$ depends on
the boundary conditions in general.
Let us take $d=5$ as an example, where
 the Weyl anomaly has only boundary contributions
\begin{eqnarray}
  \label{Ageneral}
\mathcal{A}=\int_{\partial M} \sqrt{h} [ b_1 F_{na}F^{na}+b_2 F_{ab}F^{ab}]+... \, .
\end{eqnarray}
Here $b_1, b_2$ are boundary central charges which depends on the
choices of boundary 
conditions. 
By using our key formula (\ref{key}) together with 
$A_b=a_b -x F_{bn}+ \cdots $
and the gauge choice $A_x=0$, we obtain
\begin{eqnarray}
  \label{5dcase}
\alpha_1=-2 b_1, \quad \a_2 =4 b_2,
\end{eqnarray}
which implies that the current
\eq{currenthigherd}
depends on boundary conditions for $d=5$.
Note that the first relation in \eq{5dcase} for $\a_1$ actually holds
for general curve space. 
For the free complex scalar theory,
the coefficient $\alpha_1$ for has been derived \cite{McAvity:1990we}: 
\begin{eqnarray}\label{currentscalar}
\alpha_1=\begin{cases}
-\frac{2\Gamma[\frac{d}{2}]}{(4\pi)^{\frac{d}{2}}(d-1)} ,
\ \ \ \ \ \ \ \ \ \ \text{Dirichlet BC}\\
-\frac{((5-d) d-2) \Gamma
  \left(\frac{d}{2}-1\right)}{(4\pi)^{\frac{d}{2}}(d-3) (d-1)},
\ \text{Robin BC} .
\end{cases}
\end{eqnarray}
As we have seen, for $d>4$ the current takes different values for
Dirichlet BC and
Robin BC, which agrees with our result above.

For lower dimensions $d<4$, similar analysis as (\ref{current0}) --
(\ref{solnJ1}) gives near a plane boundary the asymptotic current density
\begin{eqnarray}
  \label{current2d3d}
J_b =
\begin{cases}
 \alpha \frac{ e^2 c}{\hbar}  F_{n b}\ x,& d=2,\\
\alpha \frac{ e^2 c}{\hbar}  F_{nb} (\beta+ n_R \ln x),& d=3,
\end{cases}
\end{eqnarray}
 where $\alpha,\beta, n_R$ are constant parameters of order
 one.
 We emphasis, however, for lower dimensions $d<4$,
 the current \eq{current2d3d} are not related to the Weyl anomaly.
 Hence
 the parameters in \eq{current2d3d}
 are not related to the central charge of the theory,
 but they are determined by the specific
 details of the
 theory. For example for free
 complex scalars, we have
$n_R=1$ for Robin BC  and $n_R=0$ for  Dirichlet BC
 \cite{McAvity:1990we}.



\begin{thebibliography}{99}
\bibitem{note1}
Below we will
spell out explicitly up to what distance scale the result holds.
\bibitem{note2}
  One should not mix this up with the quantum anomalous Hall effect which refers to
  a quantized Hall effect realized in system without external magnetic field.
\end{thebibliography}
\end{document}